\documentstyle{l-aa}
\input{psfig.sty}
\newcommand{\etal}{et al. }
\newcommand{\msol}{M$_{\odot}$}

\begin{document}
\thesaurus{01
          (11.09.1: NGC$\thinspace$5907;
           11.08.1;
	   11.16.1;
           11.19.5;
           11.19.6;
           12.04.1)}
\title{V- and I-band observations of the halo of NGC$\thinspace$5907\thanks
{Based on data obtained with the CFHT in Hawaii}}
\label{}
\author{J. Lequeux\inst{1}\and
        B. Fort\inst{1}\and
        M. Dantel-Fort\inst{1}\and
        J.-C. Cuillandre\inst{2}\and
	Y. Mellier\inst{3,2}}
\offprints{J. Lequeux
           lequeux@mesioa.obspm.fr}
\institute{Observatoire de Paris, 61 Av. de l'Observatoire, 75014 Paris, France
\and
	Observatoire Midi-Pyr\'en\'ees, 14 Av. Edouard Belin, 31400 Toulouse,
        France
\and
        Institut d'Astrophysique, 98 bis Bd. Arago, 75014 Paris, France}
\date{Received April 1996 / Accepted June 1996}
\maketitle

\begin{abstract}
Using long exposures taken in the V and I bands with a 4096$\times$4096 pixel
CCD array at the prime focus of the Canada-France-Hawaii telescope, we have
confirmed the discovery by Sackett et al. (1994) of a flat, faint luminous halo 
around the edge-on Sc galaxy NGC$\thinspace$5907. This halo is 
redder than the disk. Its nature is still an observational challenge. 
We suggest however that its color can only be accounted for by stars
less massive than 0.8$\thinspace$\msol. Their total mass in 
NGC$\thinspace$5907 is
very uncertain, but it might possibly account for the excess dynamical mass in 
this galaxy. Alternatively their distribution might trace the dark potential 
without contributing significantly to the mass. 
We give upper limits to the extinction by an 
hypothetical dust in the halo.

\end{abstract}

\section{Introduction}

The presence of dark matter in spiral galaxies has long been inferred from their
flat rotation curves. Amongst the possible candidates for this dark matter,
low-mass stars and brown dwarfs are seriously considered. While brown dwarfs are
almost invisible optically, there is a chance that the most massive amongst
the low-mass stars can be detected directly. Recently, Sackett \etal (1994),
hereafter SMHB, have found that the edge-on Sc galaxy NGC$\thinspace$5907, 
which has the
usual problem of dark matter (Sancisi \& van Albada 1987; Barnaby \& Thronson
1994, hereafter BT), is surrounded by a faint luminous halo. This halo might be
rather flat since it was detected only to 6$\thinspace$kpc from the plane of 
the galaxy. 
Very recently, James \& Casali (1996), hereafter JC, have reported a surface 
brightness gradient perpendicular to the disk of the galaxy
in the J and K bands. We discuss here the results of
very deep exposures of NGC$\thinspace$5907 in the V and I
bands at the prime focus of the Canada-France-Hawaii telescope (CFHT)
with the wide-field camera MOCAM. Section
2 describes these observations and their reduction, Section 3 presents the
results and Section 4 is a discussion. Another product of our observations is 
an upper limit to the amount of dust around NGC$\thinspace$5907, presented in 
the Appendix.

\section{Observations and reductions}

The observations were made with MOCAM, a mosaic of four 2048$\times$2048 pixel
anti-blooming CCDs at the prime focus of the CFHT, with a field of view of
14$\arcmin$$\times$14$\arcmin$ at a scale of 0$''\!\!.$206 per pixel
(Cuillandre et al. 1996). 
5$\times$30-m V frames and 9$\times$20-m I$_{\rm Cousins}$ frames 
were obtained in good photometric conditions, totalling
exposure times of 2$\thinspace$h$\thinspace$30$\thinspace$m 
and 3$\thinspace$h respectively. The frames were taken with slight
shifts with respect to each other to allow elimination of the spurious events
during the reduction of the data through a median stacking of the images.
The image quality is better than 1$\arcsec$ for the 
final V frame and 0$''\!\!.$9 for the final I frame. 
The flat-field applied to each frame was built from the 
combination of all empty fields frames taken with MOCAM during the same
new-moon period as our observations (about 15 I and 15 V frames). 
The instrument set-up was not modified during the whole period. 
The same part of the calibration field SA$\thinspace$110 (Landolt 1992)
was first observed on each of the four CCDs, then the
whole field on the full camera, giving an internal check of the quality of the
reference stars: the photometric error is of the order of 0.03 mag. in each
band. Each CCD is different because the gain of the amplifier
in the reading chain varies slightly from chip to chip.
A small correction was secured by measuring the sky background on the edges 
of each CCD and by multiplying each CCD frame
by a factor such that all these sky backgrounds are at the same level. The
sky level measured after reduction was 
21.50$\thinspace$mag.$\thinspace$arcsec$^{-2}$
in V and 19.23$\thinspace$mag.$\thinspace$arcsec$^{-2}$ in I, in excellent 
agreement with the
mean values for photometric nights at Mauna Kea, respectively 21.5 and 19.2.

\begin{figure}
\psfig{figure={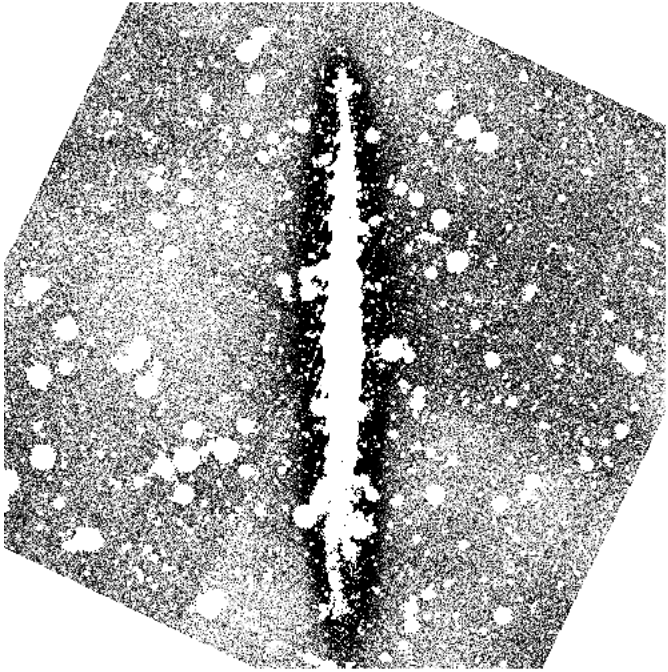},width=8.8cm}
\caption{The masked I image of NGC$\thinspace$5907. The size of the frame is 
14$'\times14'$. 
North to the upper right edge of the CCD frame, East
to the right. Note the mask applied to all bright objects, and the large-scale
irregularities of the background.}
\end{figure}
 
Several difficulties arise when one wants to reach levels of the order of 
10$^{-3}$
of the sky background. One is visible on fig.1: the background
is not fully uniform up to 3 times this level at large scales, in a way not
reproducible from frame to frame. This is due to non-uniform scattered light 
in the camera, the telescope and the dome. No correction is possible without a
dedicated effort to get a low instrumental scattered-light level as in 
coronagraphs. 
This effect is the ultimate limitation of our observations.
In this respect this situation is similar to that in the observations of
Morrison et al. (1994), hereafter MBH. 

The second effect is the existence of faint extended wings in the Point 
Spread Function (PSF),
presumably resulting from reflections between the CCD and the dewar
window, and from scattering on the optical surfaces. 
This effect is a serious limitation and we have studied
it in details. We built the PSFs in V and I to a radius of 16$''\!$ where
levels of 10$^{-7}$ of the central pixel are reached, 
using a procedure similar to that described by MBH, Sect. 3.3.4. 
Both PSFs exhibit approximate circular symmetry. Their central parts
are very well approximated by gaussians and their faint wings have a power-law
shape.  

We now discuss the different steps of data processing. A faint halo similar to
that measured by SMHB is visible on our frames, especially the I one
(fig. 1). The first step is to assess the reality of this halo, which a priori
might be due to diffused light. A simple way to check this is to 
convolve the image with the full PSF, to convolve it also with the gaussian 
central part of the PSF only, and to interpret the differences between
these two convolved images: this gives a very good approximation of the diffused
light level. We have checked in this way that long-range contamination by light
diffused from the body of the galaxy is actually negligible, as claimed by MBH.
In order to avoid contamination by the scattered
light around individual stars and galaxies in the field, we have
numerically masked their surroundings wherever the 
level of scattered light estimated from the PSF is larger than 2 times the 
r.m.s. deviation of the (unfiltered) background. 
The result is illustrated on fig. 1, where the masked
parts of the field appear blank. This procedure also eliminates the 
bright parts of the galaxy: photometry of these parts is done from the 
unmasked image. 

After eliminating the masked parts and the remaining faint stars and galaxies,
we have averaged the surface
brightness in rectangles parallel to the major axis of NGC$\thinspace$5907. The
position of the major axis has been determined from an elliptical fit of the
galaxy at relatively low brightness levels, and the galaxy image has been
rotated so that its major axis is the y axis (fig. 1). The long side of the
rectangles (along the y axis) measures 
373$\thinspace$pixels$\thinspace$=$\thinspace$1.2$\arcmin$=
$\thinspace$4.1$\thinspace$kpc and has the same length as in
SMHB in order to ease the comparison with their work (we use
the same distance of 11$\thinspace$Mpc to the galaxy). The length of the 
short side of the rectangles varies according to the 
distance to the major axis (see fig. 2). 

\section{Results}

\begin{figure}
\psfig{figure={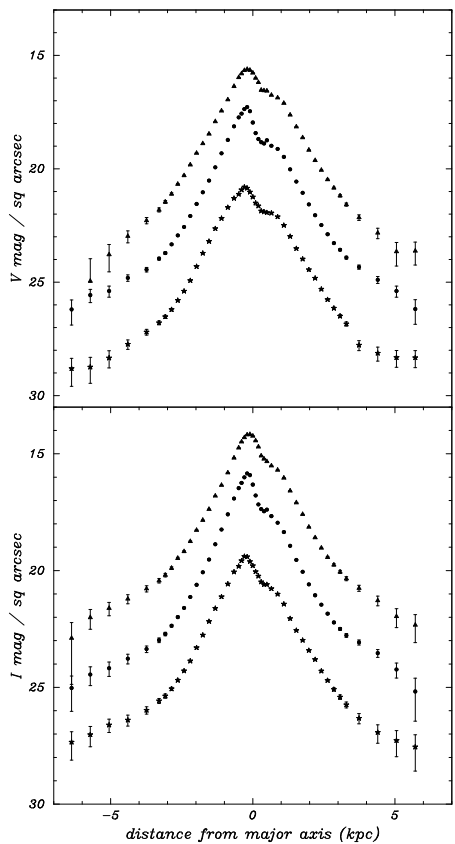},width=8.8cm}
\caption{V (top) and I (bottom) surface brightness profiles perpendicular 
to the major axis of
NGC$\thinspace$5907. The magnitude scale corresponds to the lower profile
of each set, the other
profiles being shifted upwards by 2.5 and 5 mag. respectively. Each profile 
corresponds to brightnesses averaged over 4.1$\thinspace$kpc parallel to 
the major axis.
The lower profile is centered 4.1$\thinspace$kpc below the center of the galaxy,
the middle profile at the center and the upper profile 4.1$\thinspace$kpc above 
the center. The distance to the galaxy is taken as 11.0$\thinspace$Mpc. The
errors bars are estimates of the effect of extrapolation of the space-variable
background.}
\end{figure}

\begin{figure}
\psfig{figure={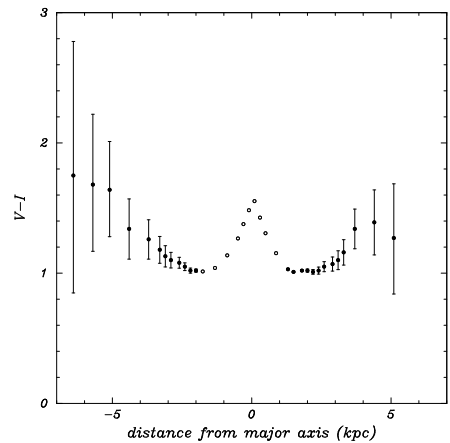},height=6cm}
\caption{The V$-$I$_{\rm Cousins}$ profile perpendicular to the major axis of
NGC$\thinspace$5907. This profile corresponds to an average over the three 
cuts of fig. 2. The errors bars result from the addition of the uncertainties 
in fig. 2}
\end{figure}

Figure 2 presents the surface brightness profiles of 
NGC$\thinspace$5907 for some of the cuts defined in the previous section. 
As explained previously the limitation is the large-scale systematic
residual flat-field variations in the background which has to be 
extrapolated under the galaxy image. 
The profiles are not completely symmetric with respect to the major axis, 
perhaps due to a warp and to the fact that the galaxy is not strictly edge-on
but has an inclination of $\approx$ 87$^{\circ}$. 

Figure 3 shows the V$-$I profile perpendicular to the plane of the galaxy, 
averaged over the three central cuts.
The central dust band appears clearly through its redder color. Aside from 
this band, the disk has V$-$I $\approx$ 1.0, a normal color for undisturbed
spirals (de Jong \& van der Kruit 1994), and by comparison with SMHB 
V$-$R = 0.65 and R$-$I = 0.35. V$-$I becomes bluer by $\approx$ 0.2 mag.
towards the edges as observed in many spirals by Peletier et al. (1995). 
However V$-$I becomes redder at increasing distances from the plane 
to reach V$-$I $\approx$ 1.35 at about 4.5$\thinspace$kpc. It seems to 
increase even further at higher z, but the data become unreliable there. 
This is true all along the disk of NGC$\thinspace$5907. At z = 
4.5$\thinspace$kpc we obtain by comparison with SMHB V$-$R $\approx$ 1.2
and R$-$I $\approx$ 0.1. We have checked by binning our original data
to the larger pixel size of SMHB and reprocessing the images that the
measurements of surface brightness in the region of interest are
essentially unaffected by the sampling.

\section{Discussion}

Our data clearly confirm the existence of a faint light emission at large 
distances from the disk of NGC$\thinspace$5907, first reported by SMHB. Is it 
due to
a luminous halo? To check this point we have built models of the light 
distribution in NGC$\thinspace$5907 based on previous work, mainly by 
van der Kruit \&
Searle (1982), BT, MBH and Fuchs (1995). The 
comparison with our observations confirms the finding of SMHB: it is 
impossible to account for the light emission above z $\approx$ 
3.3$\thinspace$kpc
without adding a halo emission. But the emission might also be possibly due 
to a faint extension
of the bulge if its ellipticity increases with radius (note that the 
distinction between bulge and halo then becomes semantic).
Another possibility is that the 
galaxy presents a very large-scale warp parallel to the plane
of the sky. However such a warp is expected to have a color V$-$I $\approx$
0.8 as the outer disk, rather than being redder than the disk. Thus
we believe that the halo emission is real and is redder than the
unabsorbed parts of the disk.  

A gradient in surface brightness corresponding to the luminous halo 
has been found independently
by JC in the J and the K band within 3.2 $\leq |z| \leq$ 6.9$\thinspace$kpc 
above the center of the galaxy. Their method does not yield a value for the 
surface brightness, but they obtain a color 
J$-$K$\thinspace$=$\thinspace$1.3$\thinspace\pm$0.3 in this
region from a comparison between the gradient in J and in K.
However our observation shows a strong color gradient in V$-$I. If 
such a gradient also exists in J$-$K,
this may affect their determination of the color. 
Assuming that the light is pure
halo light and that it is not affected by interstellar reddening 
(see appendix), it is
possible to derive some conclusions about the stellar population which is
presumably responsible for the emission. 

i) Any stellar population of any age $\leq$ 1.5 10$^{10}$yr with a ``normal''
initial mass function (hence dominated by giants) has J$-$K $\leq$ 1.0,
a value only reached for old populations 
with solar or super-solar metallicities (Buzzoni 1989). The same is true
for isolated main-sequence stars (Kirkpatrick \etal 1993, 
Allard \& Hauschildt 1995: fig. 13). 
As a high metallicity is unlikely for halo stars, we conclude that
the determination of J$-$K of JC may have problems due to the probable color 
gradient (see above).

ii) The V$-$I color of the halo may be appreciably redder than the value
of $\approx$ 1.35 we measure at 4.5$\thinspace$kpc. We have hints for redder 
colors at larger distances, and the measurements at 4.5$\thinspace$kpc may 
still be contaminated by a warp in the disk. Taking the measured color at face 
value, it is similar to that of metal-rich elliptical galaxies (see e.g. 
Poulain \& Nieto 1994). A high metallicity being unlikely for a halo, the
stellar population is rather 
dominated by stars with mass lower than 0.8 \msol, implying a cut-off
in the initial mass function since all stars with such masses are still on 
the main sequence whatever their ages. We already found a similar result
for the outer bulge of NGC$\thinspace$7814 (Lequeux \etal 1995). The visible 
luminosity 
and color of such a population is dominated by the most massive stars. For
a given color, their mass and luminosity depend much of their metallicity.
For a typical 1/100 solar halo metallicity, a V$-$I of 1.35 corresponds to 
12 Gyr-old 
stars of 0.15 \msol with M$_{\rm I}$ $\approx$ 10 (G. Chabrier, private 
communication). $\mu_{\rm I} 
\approx$ 26$\thinspace$mag.$\thinspace$arcsec$^{-2}$
corresponds to 2.5 10$^{7}\thinspace$\msol$\thinspace$kpc$^{-2}$ 
if only these stars are present. 
This is considerably lower than the 
4 10$^{8}\thinspace$\msol$\thinspace$kpc$^{-2}$
required for the minimum spherical massive halo solution of BT.
But this is a strict lower limit, and one could easily reconcile 
the masses by adding lower-mass stars which contribute little to the luminosity.
Alternatively, the faint luminous halo could just be a natural dynamical
signature of the existence of a dark halo of different nature as 
suggested by Fuchs (1995). Another possibility is that the faint light
corresponds to a flattened extension of the bulge with a de Vaucouleurs
r$^{1/4}$ law which would fit well the outer profiles of fig. 2. 
Unfortunately our measurements at large galactocentric radii are of 
insufficient quality to check this.

An interesting problem is the origin of the truncated mass function of the
very red stars. No dynamical process at the scale of galaxies can sort out 
stars by mass, because the corresponding relaxation times are much 
longer than the age of the Universe. Consequently 
the stars must have been formed with their present mass function.

\section{Conclusions} 

We have confirmed the detection by Sackett et al.
(1994) of a faint luminous halo around NGC$\thinspace$5907.
This halo is very red and can only contain stars definitely less
massive that the limit for the mass of 15 Gyr-old main-sequence stars, 
about 0.8 \msol. The actual value of their mass is unfortunately very 
dependent on their metallicity. It is not impossible, as suggested by SMHB, 
that such stars account for the missing mass in
NGC$\thinspace$5907. But the visible halo could also trace the dark matter of a
halo of different nature (Fuchs 1995).
The truncated mass function of these stars 
must be the outcome of star formation itself. This is
reminiscent of the problem of cooling flows around massive galaxies in clusters,
which if they really exist can only form low-mass stars. It could be related
to the formation of the bulge, as perhaps in NGC 7814 (Lequeux \etal
1995).

The present results are preliminary and must be confirmed by observations
using special procedures for reducing the effect of the scattered light at
very faint surface brightness levels.\\

{\bf Appendix: determination of extinction.} We used the algorithm SExtractor
(Bertin \& Arnouts 1996) to separate background galaxies from stars in both
I and V frames, and measured the V$-$I color of galaxies in the same way as 
described in Lequeux \etal (1995). We then averaged the colors on 1$'$
wide strips parallel to the major axis of NGC$\thinspace$5907. The r.m.s. 
error on the 
averaged colors, due to the intrinsic color dispersion of the galaxies, is 
about 0.025 mag. The galaxies in the central strip are redder as expected, by 
0.09 mag., a 2.8$\sigma$ effect. Those in the next strip centered on the right
side (see fig. 2) are marginally redder by 0.07 mag., at 2.5$\sigma$, 
possibly indicating dust in a warp. All the other strips have the same average
V$-$I within the uncertainties, showing that there is no significant extinction 
gradient in the observed part of the halo of NGC$\thinspace$5907. It is clear 
that the V$-$I color excess in the optical
halo cannot be greater than 0.1 mag. at most, so that it can be neglected.

\acknowledgements{We thank G. Chabrier and his collaborators, M. Casali and
P. James for interesting discussions}

\end{document}